\documentclass[aps,prd,twocolumn,superscriptaddress,amsmath,showpacs,nofootinbib]{revtex4-1}
\usepackage{graphicx}
\usepackage{dcolumn}
\usepackage{bm}
\usepackage{natbib}
\usepackage{multirow}
\usepackage{verbatim}

\newcommand{\be}{\begin{equation}}
\newcommand{\ee}{\end{equation}}
\newcommand{\bea}{\begin{eqnarray}}
\newcommand{\eea}{\end{eqnarray}}

\newcommand{\etal}{\textit{et al.}}

\urldef\mgcamb\url{http://www.sfu.ca/~aha25/MGCAMB.html}

\newcommand{\Dm}{\Delta m^2_{31}}
\newcommand{\dm}{\Delta m^2_{21}}

\newcommand{\ths}{\theta_{12}}
\newcommand{\tha}{\theta_{23}}
\newcommand{\thb}{\theta_{13}}
\newcommand{\mb}{m_\beta}
\newcommand{\mbb}{m_{\beta\beta}}

\newcommand{\thalf}{T_{1/2}^{0\nu}}

\newcommand{\Ge}{^{76}\mathrm{Ge}}
\newcommand{\Xe}{^{136}\mathrm{Xe}}

\newcommand{\meV}{\mathrm{meV}}
\newcommand{\eV}{\mathrm{eV}}
\newcommand{\keV}{\mathrm{keV}}

\newcommand{\yr}{\mathrm{yr}}

\let\oldsqrt\sqrt
\def\sqrt{\mathpalette\DHLhksqrt}
\def\DHLhksqrt#1#2{%
\setbox0=\hbox{$#1\oldsqrt{#2\,}$}\dimen0=\ht0
\advance\dimen0-0.2\ht0
\setbox2=\hbox{\vrule height\ht0 depth -\dimen0}%
{\box0\lower0.4pt\box2}}

\begin{document}
\title{The $\nu$ generation: present and future constraints on neutrino masses from global analysis of cosmology and laboratory experiments}
\author{Martina Gerbino}\email{martina.gerbino@uniroma1.it}
\affiliation{Physics Department and INFN, Universit\`a di Roma ``La Sapienza'', Ple Aldo Moro 2, 00185, Rome, Italy}

\author{Massimiliano Lattanzi}
\affiliation{Dipartimento di Fisica e Scienze della Terra, Universit\`a di Ferrara and INFN sezione di Ferrara, Polo Scientifico e Tecnologico - Edificio C Via Saragat, 1, I-44122, Ferrara, Italy}

\author{Alessandro Melchiorri}
\affiliation{Physics Department and INFN, Universit\`a di Roma ``La Sapienza'', Ple Aldo Moro 2, 00185, Rome, Italy}

\begin{abstract}
\noindent We perform a joint analysis of current data from cosmology and laboratory experiments to constrain
the neutrino mass parameters in the framework of bayesian statistics, also accounting for uncertainties in nuclear modeling,
relevant for neutrinoless double $\beta$ decay ($0\nu2\beta$) searches. We find that a combination of current oscillation,
cosmological and $0\nu2\beta$ data constrains $\mbb~<~0.045\,\eV$  ($0.014 \, \eV < \mbb < 0.066 \,\eV$) at 95\% C.L. for normal (inverted) hierarchy. 
This result is in practice dominated by the cosmological and oscillation data, so it
is not affected by uncertainties related to the interpretation of $0\nu2\beta$ data, like  nuclear modeling, or the exact particle physics mechanism 
underlying the process. We then perform forecasts for forthcoming and next-generation
experiments, and find that in the case of normal hierarchy, given a total mass of $0.1\,$ eV, and 
assuming a factor-of-two uncertainty in the modeling of the relevant nuclear matrix elements,
it will be possible to measure the total mass itself, the effective Majorana mass and the effective electron mass 
with an accuracy (at 95\% C.L.) of $0.05$, $0.015$, $0.02\,\eV$ respectively, as well as to 
be sensitive to one of the Majorana phases. This assumes that neutrinos are Majorana particles and that the mass mechanism
gives the dominant contribution to $0\nu2\beta$ decay. We argue that more precise nuclear modeling
will be crucial to improve these sensitivities.
\end{abstract}


\maketitle


\section{Introduction}
\noindent It is by now firmly established by oscillation experiments that neutrinos do have a mass.
However, oscillation experiments are only sensitive to neutrino mass differences and 
mixing angles, and thus
do not provide information on the absolute scale of masses, on the mass hierarchy
nor on their Dirac or Majorana nature.  
The nature of neutrino masses and their smallness with
respect to those of the charged leptons represents a puzzling fact, possibly
related to the mechanism of neutrino mass generation. Three main avenues are currently being pursued in order to
experimentally probe the absolute scale of neutrino masses, namely i) direct measurements, studying the kinematics of  $\beta$ 
decay \cite{Drexlin:2013lha}, ii) searches for neutrinoless double $\beta$ decay ($0\nu2\beta$) \cite{Cremonesi:2013vla}, and iii) cosmological observations \cite{Lesgourgues:2006nd}.
Approaches based on kinematic arguments
have the advantage of being very direct and model-independent. An alternative is to study $0\nu2\beta$ decay,
\emph{i.e.}, the double $\beta$ decay of nuclei, in which no neutrinos are present in the final state. 
If observed, it would guarantee that 
neutrinos have a non-vanishing Majorana mass \cite{Schechter:1981bd};
if not, upper limits on the mass scale can still be placed, under the assumption that neutrinos are Majorana particles.
Relating the (potentially) observed rate for this process to neutrino masses also requires to assume 
that the mass mechanism is the dominant one leading to $0\nu2\beta$ decay. It is worth noting that
even if this is the most natural scenario, nevertheless other possibilities exist, involving additional
physics beyond the standard model, see e.g. Refs. \cite{Deppisch:2012nb,Rodejohann:2011mu} for a discussion.
Moreover,
our imprecise knowledge of the appropriate nuclear matrix elements is a relevant source of uncertainty
on the interpretation of the results of these experiments \cite{Cremonesi:2013vla}. Finally, neutrino masses can be measured through cosmological
observations, like measurements of the temperature and polarization anisotropies of the cosmic microwave background, or of the distribution
of large scale structures, since massive neutrinos affect the background evolution of the Universe, as well as the growth of cosmological perturbations.
Cosmology presently provides the most stringent limits on the absolute scale of neutrino masses  \cite{Planck:2015xua}, with the shortcoming 
that these limits depend on assumptions on the underlying cosmological model.

The three approaches outlined above should be seen as complementary, as each of them
presents its own advantages and disadvantages, and also because they probe slightly different quantities 
related to the neutrino masses. For this reason, it appears natural to combine data from 
direct measurements, $0\nu2\beta$ searches and cosmology, other than from oscillation experiments, in order to 
constrain the neutrino mass parameters \cite{joint}. In this paper, we want to derive joint constraints on 
neutrino mass parameters from the most recent observations from both laboratory and cosmological experiments,
combining them in the framework of Bayesian statistics. In particular, for $0\nu2\beta$ experiments, 
we take into account the uncertainty related to nuclear matrix elements, by treating it as a nuisance parameter to be 
marginalized over, in order to account its impact on the neutrino mass estimates. We also perform forecasts, considering both
forthcoming and next-generation experiments.

\section{Method}

\noindent We use $m_i$ $(i=1,\,2,\,3$) to denote the masses of the neutrino mass eigenstates $\nu_i$. We denote 
with 1 and 2 the eigenstates that are closest in mass; moreover, we take $m_2 > m_1$, so that $\dm$ is always positive, while the sign of $\Dm$ discriminates between the normal (NH) and inverted (IH) hierarchies,
for $\Dm > $ or $<0$, respectively.
The neutrino mass eigenstates are related to the flavour eigenstates $\nu_\alpha$ ($\alpha=e,\,\mu,\,\tau$) through
$\nu_\alpha = \sum_i U_{\alpha i} \nu_i$, 
where $U_{\alpha i}$ are the elements of the neutrino mixing matrix $U$,
parameterized by the three mixing angles $(\ths,\,\tha,\,\thb)$, one Dirac ($\delta$) and two Majorana ($\alpha_{21},\,\alpha_{31}$) CP-violating phases.
Oscillation phenomena are
insensitive to the two Majorana phases, that however affect lepton number-violating processes
like $0\nu2\beta$ decay.
The different probes of the absolute scale of neutrino masses are sensitive to different
combinations of the mass eigenvalues and of the elements of the mixing matrix. $\beta$ decay
experiments measure the squared effective electron neutrino mass
$\mb^2 \equiv  \sum_i \left| U_{ei}\right|^2m_i^2$,
while $0\nu2\beta$ searches are sensitive to the effective Majorana mass
$\mbb\equiv \left| \sum_i U_{ei}^2 m_i\right|$,
where $\phi_2 \equiv \alpha_{21}$ and $\phi_3 \equiv \alpha_{31} - 2\delta$.
Finally, cosmological observations probe, at least in a first approximation, the sum of neutrino masses
$M_\nu \equiv \sum_i m_i = m_1+m_2+m_3$.

\noindent We perform a bayesian analysis based on a Markov chain Monte Carlo (MCMC) method,
using \texttt{cosmoMC} \cite{Lewis:2002ah} as a generic sampler in order to explore
the posterior distribution of the parameters given the data.
We consider the following vector of base parameters:
$\left(M_\nu,\,\dm, \Dm,\,\sin^2 \ths,\,\sin^2 \thb,\,\phi_1,\,\phi_2,\,\xi \right)$
where $\xi$ is a ``nuisance'' parameter related to the uncertainty in nuclear modeling (see below). We assume uniform prior distributions for all parameters. We do not consider the mixing angle $\theta_{23}$ since none of the mass parameters depend on it. 

\noindent We consider data from oscillation experiments, direct measurements of the 
electron neutrino mass, $0\nu2\beta$ searches and cosmological observations, all folded in the analysis
through the corresponding likelihood function.
Our baseline dataset is the most recent global fit of the neutrino oscillation parameters \cite{Forero:2014bxa},
updated after the Neutrino 2014 conference. We model the likelihood as a the product of individual gaussians
in each of the oscillation parameters, since correlations can be neglected for our purposes \cite{Forero:2014bxa, GonzalezGarcia:2012sz, Bergstrom:2015rba}. 
For the means and standard deviations, we take
respectively the best-fit value and the $1\sigma$ uncertainty quoted in Tab. II of Ref.~\cite{Forero:2014bxa}. When the error
is asymmetric, we conservatively take the standard deviation equal to the largest between the left and right uncertainties.
For direct measurements, we consider KATRIN \cite{Osipowicz:2001sq} and HOLMES  \cite{Alpert:2014lfa} 
as our forthcoming and next-generation datasets, 
respectively. KATRIN is expected to reach sub-eV sensitivity in $\mb$, while HOLMES 
could go down to $100\,\meV$. 
Kinematic measurements are directly sensitive to the square of the effective electron neutrino mass,
so in both cases we take the likelihood to be a gaussian in $\mb^2$ (with the additional condition that $\mb^2\ge 0$),
with a width given by the expected sensitivity of the experiment, i.e. $\sigma(\mb^2) = 0.025,\, 0.006\,\eV^2$ for KATRIN and HOLMES, respectively.
For $0\nu2\beta$ searches, we consider the current data from the GERDA experiment \cite{Agostini:2013mzu} as the present dataset,
its upgrade to the so-called ``phase 2'' for the near-future, and the nEXO experiment \cite{nEXO} as a next-generation dataset. 
$0\nu2\beta$ experiments are sensitive to the half-life of $0\nu2\beta$ decay $\thalf$. 
Assuming the Majorana nature of neutrinos, and that $0\nu2\beta$ decay is induced by the exchange of light Majorana neutrinos (in the following
we shall always assume that this is the case, unless otherwise stated),
$\thalf$ is  related to the Majorana effective mass through:
\begin{equation}
\thalf = \frac{1}{G^{0\nu}\left|\mathcal{M}^{0\nu}\right|^2}\frac{m_e^2}{\mbb^2}
\end{equation}
where $m_e$ is the electron mass, $G^{0\nu}$ is a phase space factor and $M^{0\nu}$ is the nuclear matrix element. 
The phase I of the GERDA project provides the tightest bounds on the half-life of $0\nu2\beta$ decay of $\Ge$, reporting a limit $\thalf > 2.1 \times 10^{25}\,\yr$
at 90\% C.L. ($\mbb < 200 - 600\,\meV$) \cite{Agostini:2013mzu}\footnote{
Other isotopes currently yield $\thalf > 4.0 \times 10^{24}\,\yr$ ($^{130}\mathrm{Te}$) \cite{Alfonso:2015wka} and $\thalf > 2.6 \times 10^{25}\,\yr$ ($\Xe$) at 90\% C.L. \cite{Asakura:2014lma},
corresponding to $\mbb < 270 - 760\,\meV$ and $\mbb < 140 - 280\,\meV$, respectively.
}. The upgrade to the phase II of the experimental program is expected to increase
the 90\% C.L. sensitivity to $\thalf > 1.5\times10^{26}\,\yr$, ($\mbb < 90 - 150 \, \meV$) for 40 kg of detector mass and 3 years of observations \cite{NOWcatta}.
nEXO is a next-generation ton-scale experiment for the detection of $0\nu2\beta$ decay of $\Xe$, conceived as scaled-up version of the currently
ongoing project EXO, with an estimated sensitivity $\thalf > 6.6\times 10^{27}\,\yr$ at 90\% C.L. ($\mbb < 7 - 18\,\meV$) for 5 tons of material and 5 years of data \cite{NOWpocar}.
We model the likelihood of $0\nu2\beta$ experiments as a Poisson distribution in the number of observed events in the ``region of interest'' 
(the energy window around the $Q$-value of the decay) with an 
expected value $\lambda = \lambda_S + \lambda_B$ given by the sum of signal ($S$) and background  ($B$) contributions. For a given value of $\thalf$, the expected number of signal events observed in a time $T_\mathrm{obs}$ for a detector mass $M$ is
\begin{equation}
\lambda_S=\frac{\ln 2 N_A \mathcal{E} \epsilon}{m_{enr} \thalf}\, ,
\end{equation}
where $N_A$ is Avogadro's number, $\mathcal{E}\equiv M T_\mathrm{obs}$ is the exposure, $\epsilon$ is the detector efficiency, $m_{enr}$ is the molar mass of the enriched element involved in the decay. The level of background is usually expressed in terms of the ``background index'', i.e. the number
of expected background events per unit mass and time within an energy bin of unit width. For GERDA-I, we use the parameters 
reported in Tab. I of \cite{Agostini:2013mzu} for the case with pulse-shape discrimination. For GERDA-II, we consider
a reduction of the background index down to $10^{-3}\,\mathrm{counts}\,\keV^{-1}\mathrm{kg}^{-1}\yr^{-1}$, 
a total exposure of 120 kg yr, and the same efficiency as GERDA-I \cite{priv_catta}. For nEXO, we assume a background index corresponding 
to 3.7 events $\mathrm{ton}^{-1}\yr^{-1}$ in the region of interest and an exposure of 25 ton yr \cite{NOWpocar}, and the same
efficiency as EXO \cite{Albert:2014awa}. We also consider an update to nEXO in which the background in the inner 3 tons of the detector can be 
reduced by a factor 4 through $\mathrm{Ba}$ tagging. We assume 10 years of observations for this updated version \cite{NOWpocar}.

In order to account for the uncertainty related to nuclear modeling \cite{nuclear},
including both that on nuclear matrix elements (NME) and that on the axial coupling constant, 
we compute $\thalf$ for a given $\mbb$ using fiducial values of these quantities, and then rescale it
by a factor $\xi^2$. A similar approach was used in Ref. \cite{Minakata:2014jba} in a frequentist framework, while we refer to Ref. \cite{Bergstrom:2012nx} for a different bayesian approach. The fiducial values are $g_A = 1.273$ for the axial coupling, $G^{0\nu} = 2.363\times10^{-15} \yr^{-1}$ ($14.58 \times 10^{-15} \yr^{-1}$) and $\mathcal{M}^{0\nu}=3$ ($2$) for $\Ge$ ($\Xe$).
The value of $\xi$ is extracted at every step of the MC from a uniform distribution
in the range $[0.5,\,2]$, and marginalized over. This is equivalent, for example, to assume 
that, given exact knowledge of the axial coupling, the numerical estimates
of the NME can be wrong by up to a factor two in either direction. Finally, for what concerns the cosmological dataset,
we use results obtained combining full mission Planck temperature and polarization data with data on the baryon acoustic oscillations \cite{Planck:2015xua},
as both our current and forthcoming reference dataset. For simplicity, we shall refer to this dataset simply as ``Planck 2015''.
In particular, we use the chains publicly available through the Planck Legacy Archive \cite{PLA} to derive the posterior distribution of $M_\nu$
given these data, corresponding to a 95\% upper limit $M_\nu < 0.17\,\eV$. As a next-generation experiment,
we consider the Euclid mission. The combination of all Euclid probes (weak lensing tomography, galaxy clustering and ISW) with data from Planck is expected to constrain the sum of neutrino masses
with a sensitivity of $0.06\,\eV$ for $M_\nu = 0.1\,\eV$, as reported in Tab.2 and the main text in \cite{Laureijs:2011gra}. We shall refer to this dataset simply as ``Euclid''.
We model the likelihood as gaussian in $M_\nu$, with
$\sigma(M_\nu)=0.06 \, \eV$ and the addition of the physical prior $M_\nu >0$. 

To summarize, we consider four combinations of datasets. All of them include the most updated information from oscillation experiments. 
The ``present'' dataset includes Planck 2015 for cosmology, and GERDA-I for $0\nu2\beta$ searches
We do not include information from available 
direct measurements (e.g. those from the Troisk and Mainz experiments) since they do not add information on
$\mb$ with respect to the data already considered. 
The ``forthcoming'' dataset consists of the same cosmological data as the previous dataset,
GERDA-II, and KATRIN for kinematic measurements. The ``next generation I (II)'' dataset includes Euclid, nEXO without (with)
$\mathrm{Ba}$ tagging and HOLMES. For future data, we have to assume fiducial values of the parameters: in the case of the forthcoming dataset,
we take them equal to their best estimates from the combination of oscillations and Planck2015. For the futuristic case,
we assume $M_\nu = 0.1\,\eV$ and estimate $\mb$ and $\mbb$ from the combination of Euclid and oscillation parameters. 

\section{Results} 
\begin{figure*}[htb!]
\begin{center}
\begin{tabular}{c c c}
\includegraphics[width=0.33\textwidth]{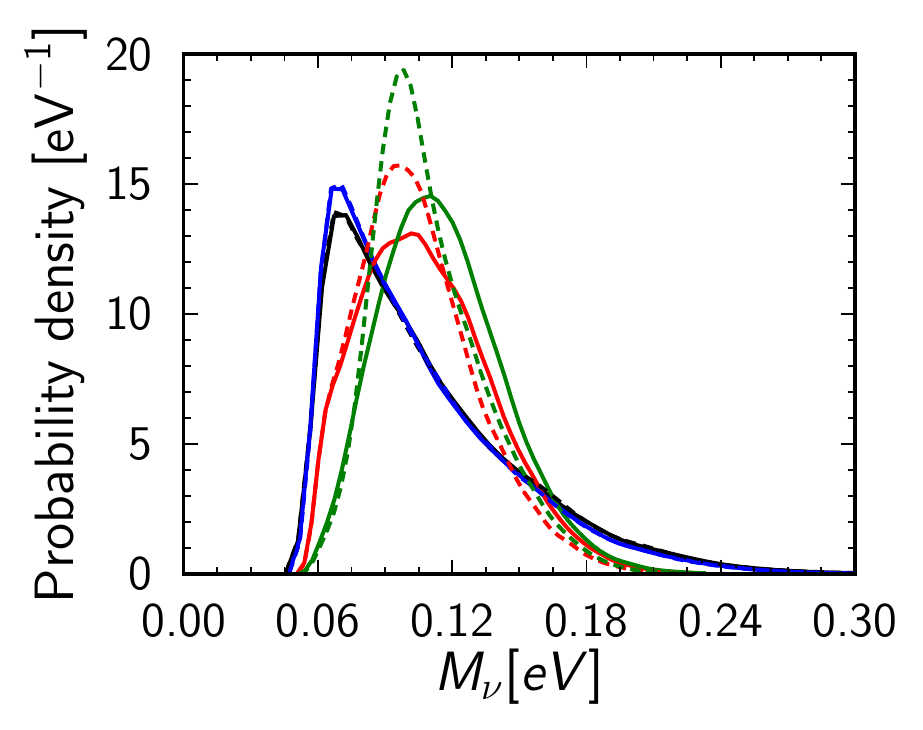} & 
\includegraphics[width=0.33\textwidth]{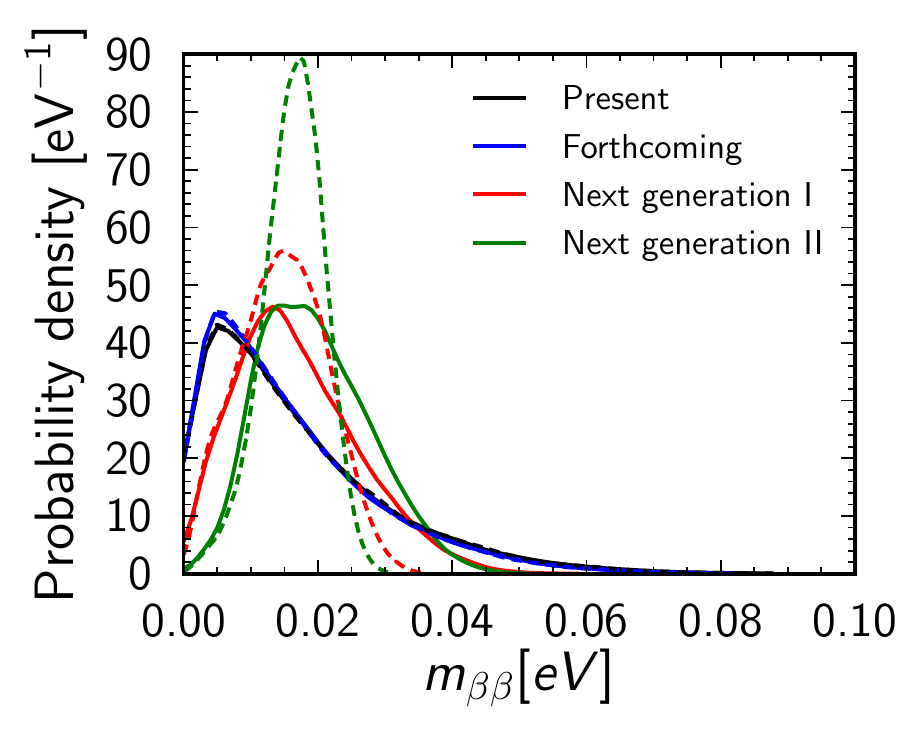} &
\includegraphics[width=0.33\textwidth]{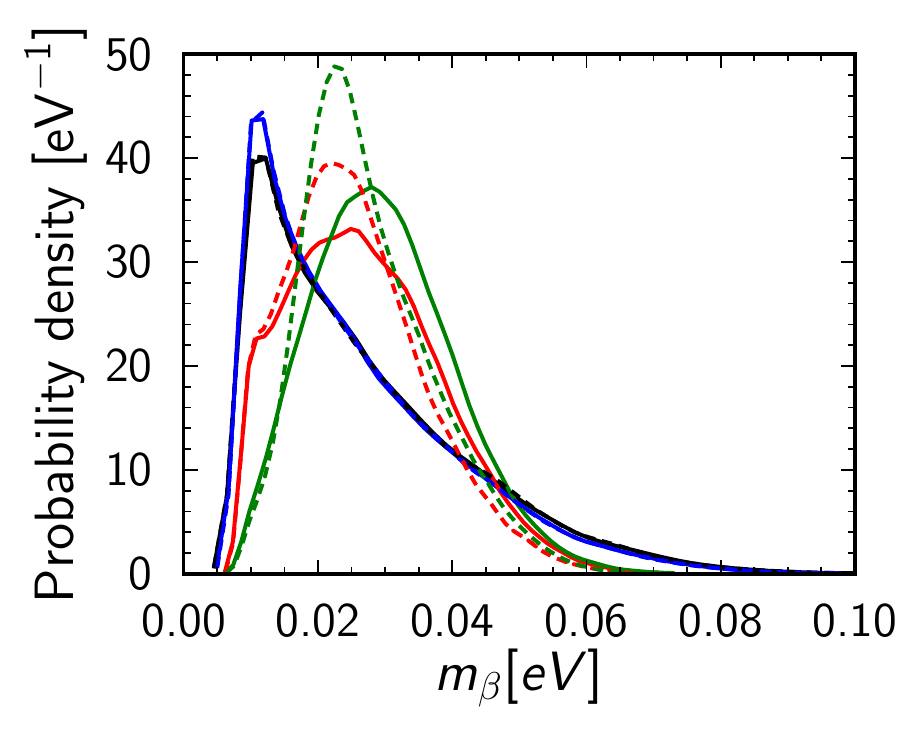} \\
\includegraphics[width=0.33\textwidth]{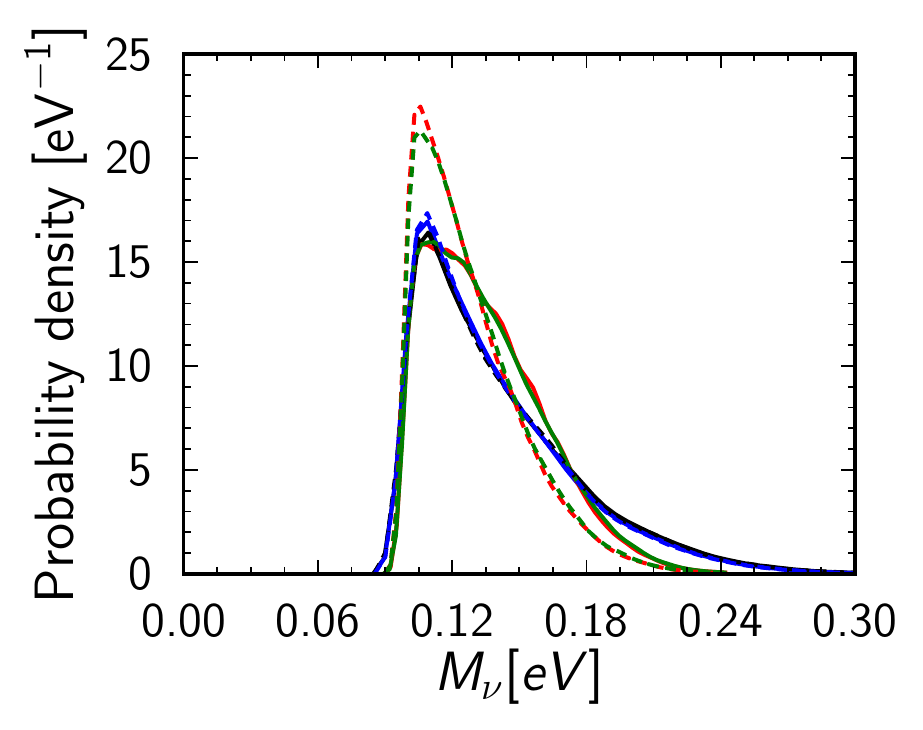} & 
\includegraphics[width=0.33\textwidth]{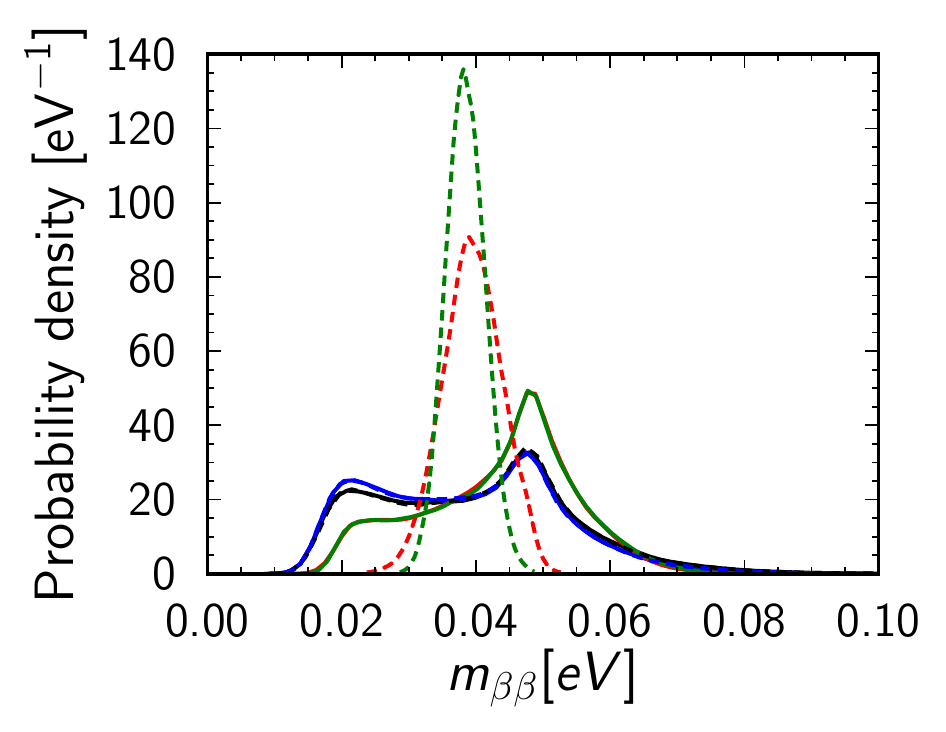} &
\includegraphics[width=0.33\textwidth]{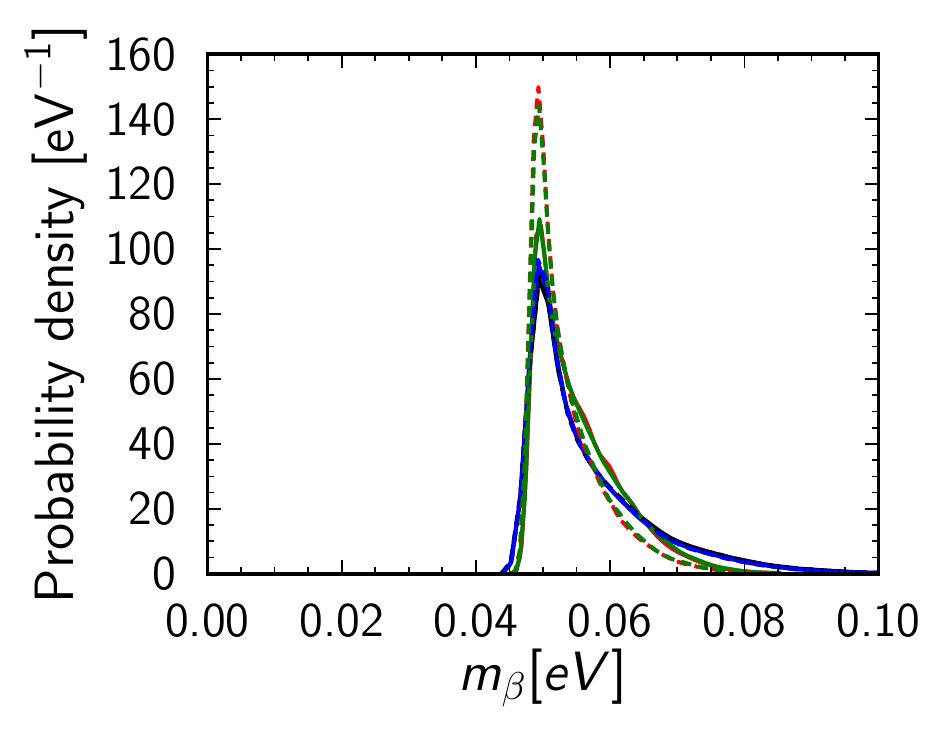} 
\end{tabular}
\end{center}
\caption{Posterior distributions for the neutrino mass parameters, for NH (top row) and IH (bottom row). 
Solid (dashed) curves correspond to marginalization over nuclear uncertainties (fixed fiducial values for nuclear parameters). Note that the ``Next-generation I'' and ``Next-generation II'' forecasted results have been derived by assuming a fiducial value of $M_{\nu}=(0.10\pm0.06)\eV$.}
\label{fig:post1D}
\end{figure*}

\begin{figure}[htb!]
\begin{center}
\includegraphics[width=0.7\columnwidth]{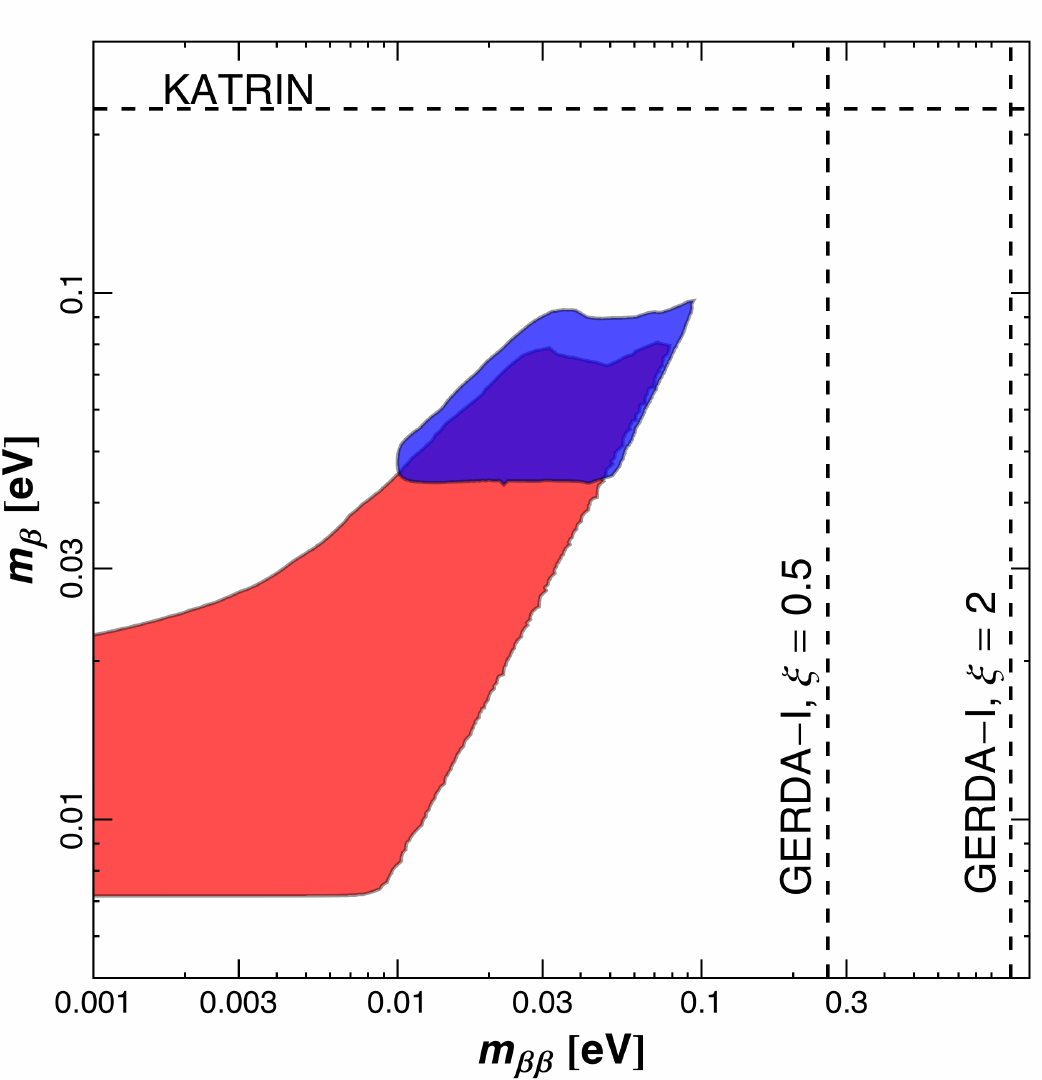}
\end{center}
\caption{Two-dimensional posterior distributions for the neutrino mass parameters in the $m_{\beta\beta}-m_{\beta}$ plane, for NH (red) and IH (blue), from the combination of oscillation and ``Planck 2015'' datasets. Contours correspond to 95\% C.L.. The dashed lines are the 95\% C.L. upper limits on $\mbb$ from GERDA phase 1 within the range $\xi=[0.5-2]$ (vertical) and on $\mb$ from KATRIN (horizontal).}
\label{fig:post2D}
\end{figure}

\begin{figure}[htb!]
\begin{center}
\includegraphics[width=0.8\columnwidth]{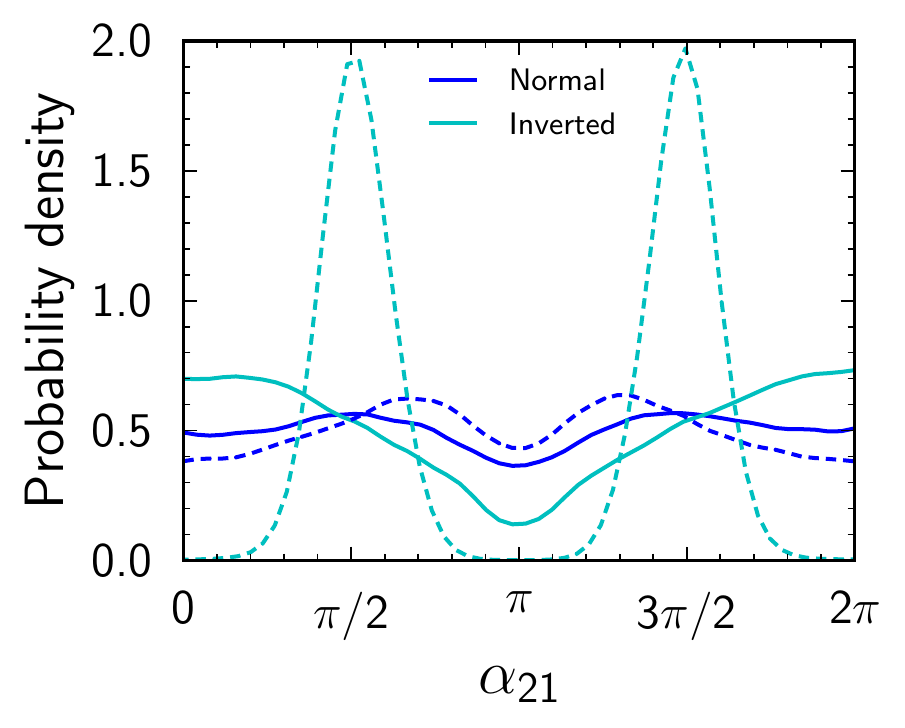}
\end{center}
\caption{Posterior distribution for $\alpha_{21}$ from the next-generation II dataset. Solid (dashed) lines
are for $\xi$ marginalized over ($\xi=1$).}
\label{fig:phi2}
\end{figure}

\noindent We present our results for $M_\nu$, $\mb$ and $\mbb$ in Tab. \ref{res1} for the three datasets described above.
We report limits both in the case where $\xi$ is fixed to 1, and when $\xi$ is marginalized over, in order to show the
impact of uncertainties in nuclear modeling. We quote our results, both in the text and table,
in terms of the Bayesian 95\% minimum credible interval  \cite{Hamann:2007pi}. When this interval
includes the minimal value of the parameter allowed by oscillation measurements, we only quote
the extremes of the range; on the contrary, we report the mean $\pm$ the 95\% uncertainty. We do this in order to emphasize a ``detection'' scenario -- i.e., one in which
the observations point to a value of the parameter under consideration being different, with a given statistical significance, from
the lowest value allowed by oscillations alone -- from a ``non-detection'' scenario in which this oscillation minimal value is still allowed.
We choose to identify the minimal value allowed by oscillations as the bayesian 95\% C.L. lower limit of the neutrino mass parameters when the lightest eigenstate is set to zero. With this definition, we get $M_\nu^{\mathrm{min}}=0.057,\eV$ ($M_\nu^{\mathrm{min}}=0.096\,\eV$), $\mb^{\mathrm{min}}=0.009\,\eV$ ($\mb^{\mathrm{min}}=0.047\,\eV$), $\mbb^{\mathrm{min}}=0.002\,\eV$ ($\mbb^{\mathrm{min}}= 0.016\,\eV$) for NH (IH). We would like to point out that the exact definition of the minimal value is somehow arbitrary, in a sense that it is not formally well-defined due to the finite precision of the oscillation measurements. For example, we could have chosen the lowest value allowed by fixing the oscillation parameters to their best-fit values, rather than computing the bayesian 95\% lower limit, and this would equally make sense. This choice only affects the way in which limits are reported in Tab. \ref{res1} (and we verified that it also has a minor impact in that respect), so it does not alter our conclusions in any way. 
In any case, we recall that the confidence intervals represent a compression of the information contained in the one-dimensional posteriors,
that fully represent the probability distribution associated to a given parameter.
In Fig. \ref{fig:post1D} we show
the marginalized one-dimensional posterior distributions for the mass parameters. In most cases, the low mass region
is excluded by the oscillation data, with the only exception of $\mbb$ in the case of NH; the reason
is that in this case the phases can arrange in order to yield $\mbb=0$ even for finite values of the
mass differences. Present data provide similar limits
independently of whether nuclear uncertainties are marginalized over. This happens because the present constraints 
are dominated by the cosmological limit on $M_\nu$, that translates directly to bounds on $\mb$ and $\mbb$ once
oscillation data are taken into account (this can be understood by noticing that the direct limits on these parameters 
are much weaker). We have verified explicitly that this is the case by performing parameter estimation using only Planck2015 and oscillation data, as shown in Fig.\ref{fig:post2D}.
In particular, we find that the present data  
constrain $\mbb~<~0.045\,\eV$ ($0.014\,\eV < \mbb < 0.066\,\eV$) at 95\% C.L. for NH (IH), regardless
of the inclusion of $0\nu2\beta$ information.
Forthcoming datasets yield similar constraints for the mass parameters;
this means that the improved sensitivity of GERDA-II and
the inclusion of KATRIN add only marginally to the Planck2015 plus oscillations data combination.
The fact that present and forthcoming limits on $\mbb$ are dominated by the latter dataset has the consequence
that they do not depend on the modeling of $0\nu2\beta$-decaying nuclei, nor on assumptions about the mechanism that induces the decay
(while, on the other hand, they are affected by the model dependence of the cosmological analysis).
This picture changes substantially for next-generation experiments. In this case, cosmological observations
and $0\nu2\beta$ searches have comparable power in constraining the mass parameters, and
the nuclear uncertainties  - as well as theoretical assumptions about the particle physics of 0$\nu$2$\beta$ decay -  
play a role in deriving parameter constraints.
We find that, if neutrinos are Majorana and 0$\nu$2$\beta$ decay is dominantly induced by the mass mechanism,
marginal 95\% evidence for non-minimal mass parameters can be obtained in the case of normal hierarchy,
even when nuclear uncertainties are taken into account. This detection is further strengthened
in the Next generation-II dataset, for which we get $M_\nu = 116^{+54}_{-51}\,\meV$, $\mbb = 20^{+17}_{-15}\,\meV$, $\mb = 30\pm20\,\meV$. In the case of inverted hierarchy, we obtain upper limits for $M_\nu$ and $\mb$, 
and a more than 95\% evidence for non-minimal $\mbb$. In particular, for the Next generation-II dataset and
marginalizing over nuclear uncertainties, we find $M_\nu < 187\,\meV$, $\mbb = 43^{+20}_{-24}\,\meV$ and $\mb < 69 \,\meV$.
Finally, we report that while present and forthcoming experiments have little, if none, sensitivity 
to the neutrino mixing phases, the combination of next-generation experiments will possibly allow 
to determine the value of $\alpha_{21}$, as shown in Fig. \ref{fig:phi2}.

\section{Conclusions} 
\noindent The combination of current and forthcoming data from oscillation, kinematic, $0\nu2\beta$ and cosmological experiments
allows to put upper bounds $M_\nu < 0.19 \,(0.21)\,\eV$, $\mbb < 0.04 \, (0.06) \,\eV$ and $\mb < 0.06 \,(0.08) \,\eV$ for NH (IH).
These limits are dominated by the combination of oscillations and cosmological data and as such
are not affected by uncertainties in nuclear modeling, nor rely on the knowledge
of the particle physics mechanism leading to  $0\nu2\beta$ decay. If neutrinos are Majorana particles
and $0\nu2\beta$ decay is induced by the exchange of light Majorana neutrinos, and further assuming a total mass of $0.1\,\eV$ and a 
factor 2 uncertainty in nuclear modeling, next-generation experiments will ideally allow to
measure non-minimal mass parameters with a 95\% accuracy better than $0.05$, $0.015$, $0.02\,\eV$
for $M_\nu$, $\mbb$, $\mb$ respectively, for NH. In the case of IH, the allowed parameter 
range is reduced by roughly 25\% with respect to the present for $M_\nu$ and $\mb$, while 
$\mbb$ can be measured with a $0.02\,\eV$ accuracy. The uncertainty on $\mbb$ can be reduced
by up to a factor 4 by a better modeling of the nuclear factors. Next-generation experiments will also be
sensitive to the phase $\alpha_{21}$. 

\begin{table*}[h!]
\centering
\begin{tabular}{ccccccccc}
\hline
\hline
Parameter&	\multicolumn{2}{c}{Present}&			\multicolumn{2}{c}{Forthcoming}&			\multicolumn{2}{c}{Next-generation}& \multicolumn{2}{c}{Next-generation II}\\
& Normal & Inverted & Normal & Inverted & Normal & Inverted & Normal & Inverted\\
\hline
\hline
$M_{\nu},\xi\,\mathrm{free}$ & 	$ [50-194] $& 	$[91-217]$ &		$[50-188]$& 		$[91-213]$&	$109_{-51}^{+56}$&		$[95-186]$&		$116_{-51}^{+54}$&		$[95-187]$\\
$M_{\nu},\xi\equiv1$& 	$ [50-195]$& 		$[91-217]$&		$[50-188]$& 	$[91-212]$&	$105_{-47}^{+52}$&		$[94-176]$&		$110_{-43}^{+53}$&		$[94-177]$\\
\hline
$m_{\beta\beta},\xi\,\mathrm{free}$&		$ < 45 $& 		$ [14-66]$& $<42$ & 	$[14-64]$&	$[1-35]$&		$43^{+19}_{-24}$&		$20_{-15}^{+17}$&		$43_{-24}^{+20}$\\
$m_{\beta\beta},\xi\equiv1$&		$ <45 $& 		$ [14-66]$&		$<42$& 	$[14-63]$&	$15^{+12}_{-13}$&		$39\pm9$&		$16\pm9$&		$38\pm6$\\
\hline
$m_{\beta},\xi\,\mathrm{free}$&	 	$ [5-59]$& 		$[45-78]$&		$[5-57]$& $[45-76]$&	$[8-48]$&		$[47-68]$&		$30\pm20$&		$[47-69]$\\
$m_{\beta},\xi\equiv1$&	$ [5-59]$& 	 	$[45-78]$&		$[5-57]$& 		$[45-76]$&	$[8-45]$&		$[46-65]$&	$28_{-17}^{+20}$&		$[46-66]$\\\hline
\hline
\end{tabular}
\caption{Limits on neutrino mass parameters from different datasets. For each parameter, we quote the 95\% bayesian minimum credible interval. When this interval includes the minimal value for the parameters computed as the lower limit at 95\% C.L. allowed by oscillation data, we only quote the extremes of the range; otherwise, we quote the 95\% range around the mean value. Units are in meV. Note that the ``Next-generation'' and ``Next-generation II'' forecasted results have been derived by assuming a fiducial value for the sum of the neutrino masses of $M_{\nu}=0.10\,\eV$.}
\label{res1}
\end{table*}

\begin{acknowledgments}
\noindent We would like to thank C. Cattadori, S. Dodelson, M. Hirsch, M. T\'ortola and F. Vissani for useful discussion.
MG and AM acknowledge support by the research grant Theoretical Astroparticle Physics number 2012CPPYP7 under the program PRIN 2012 funded by MIUR and by TASP, iniziativa specifica INFN.
\end{acknowledgments}

\end{document}